\documentclass[a4paper,12pt]{article}
\usepackage[utf8]{in
    putenc}
\usepackage{cancel}
\usepackage{ulem}
\usepackage{amsfonts}
\usepackage{amssymb}
\usepackage{graphicx}
\usepackage{amsmath}
\usepackage{enumerate}

\usepackage{subfloat}
\usepackage{subfig}
\usepackage{color}
\usepackage{float}
\usepackage{here}
\usepackage{cite}
\usepackage{mathrsfs}
\usepackage{float,epsfig}
\usepackage{dcolumn}

\usepackage{graphicx}
\usepackage{bm}
\usepackage{amsmath,amssymb,amsthm}
\usepackage[colorlinks=true,linkcolor=blue]{hyperref}
\title{\bf \Large 
Joule-Thomson Expansion of RN-AdS Black Holes in $f(R)$ gravity
}

\author{M. Chabab$^{1}$\footnote{mchabab@uca.ac.ma }, H. El Moumni$^{1,2}$\footnote{hasan.elmoumni@edu.uca.ma (Corresponding author)}, S. Iraoui$^{1}$\footnote{s.iraoui@edu.uca.ma}, K. Masmar$^{1}$\footnote{karima.masmar@edu.uca.ma}, S. Zhizeh$^1$\footnote{sara.zhizeh@edu.uca.ac.ma} \\
\\ {\small $^{1}$ High Energy and Astrophysics Laboratory, FSSM, P.O.B. 2390, 
    }\\
    {\small Cadi Ayyad University, 40000 Marrakesh, Morocco.
    }\\
    {\small $^{2}$ EPTHE, Physics Department, Faculty of Sciences,  Ibn Zohr University, Agadir, Morocco. }
}

\vspace{-3.6em}

\date{}
\begin{document} \maketitle
\begin{abstract}
{\noindent}

In this paper, we study Joule-Thomson expansion for charged AdS black holes in $f(R)$ gravity. We obtain the inversion
temperatures as well as inversion curves, and  investigate similarities and differences between van der Waals  fluids
and charged AdS black holes in $f(R)$ gravity  for this  expansion.
In addition, we determine the position of the inversion point versus different values of
the mass $M$, the charge $Q$ and the parameter $b$ for such black hole. 
At this point,  the Joule-Thomson coefficient $\mu$ vanishes, an import feature that we used to obtain the cooling-heating regions by scrutinizing the sign of the $\mu$ quantity.

\end{abstract}
\newpage
\tableofcontents

\section{Introduction}

Nowadays, an $f(R)$ gravity is one of important class of the modified  Einstein's gravity. In general, It is built through adding higher powers of the scalar curvature $R$, the Riemann and Ricci tensors, or their derivatives to the lagrangian description \cite{FR1,FR2,FR3,FR4,FR5}. This kind of gravity mimics successfully the history of universe, especially the  current cosmic acceleration,  the inflation and structure formation in the early Universe \cite{FR2,FR3}, various extensions of f(R) gravity theory has been elaborated ranging from three dimensional \cite{Hendi:2014mba} and asymptotically Lifshitz black hole solutions \cite{Hendi:2013zba} to F(R) gravity's rainbow model \cite{Hendi:2016oxk}.

Recently, a special attention has been devoted to the study of the black holes phase transition particularly after the introduction of the notion of the extended phase space via the identification of the cosmological constant with the pressure  and its conjugate quantity  with thermodynamic volume \cite{Kastor}. In this context, the black hole behaves like Van der Waals fluid\cite{KM,our,Hendi:2016vux} leading to a remarkable correspondence between the thermal physics of black holes and simple substances \cite{Kubiznak:2014zwa}. Numerous extensions  of these works have been elaborated for rotating  and hairy black hole  \cite{our1,our2}, high curvature theories of gravity and M-theory \cite{our3,our4,our5,our6}.
More exotic results as holographic heat engine \cite{our7} as well as other technics ranging from the behaviour of the quasi-normal modes \cite{our8,our9} to AdS/CFT tools  \cite{holo1,holo2} and chaos structure \cite{Chabab:2018lzf} have consolidated the similarity with the Van der Waals fluid.

Meanwhile, a considerable effort has been dedicated to explore the thermodynamics of the AdS black hole in the $R+f(R)$ gravity background with a constant curvature \cite{fBH6} where the essential of thermodynamical quantities like the entropy, heat capacity and the Helmholtz free energy are calculated, then the extended phase space and the critical Van der Waals-like behavior introduced in \cite{Chen:2013ce,fBH6} as well as the canonical ensembles in \cite{Li:2016wzx}. 
Here we would like to go further in the thermodynamical investigation and study the Joule-Thomson expansion for the charged-AdS black hole configuration in the $f(R)$ gravity background. 

 More recently, the authors of  \cite{jt1} have investigated  
 the {\tt JT} expansion for AdS charged black holes with the aim to confront the resulting features with those of Van der Waals fluids.  The extension to rotating-AdS black hole \cite{jt2} and the charged black hole solution in the presence of the quintessence field \cite{jt3} have also been considered. The JT expansion of a gas generally takes place at a constant enthalpy,  a quantity identified to a mass when the thermodynamical system is a black hole,  more precisely its mass. {\tt JT} expansion \cite{ref49} is a convenient isoenthalpic tool that a thermal system exhibits with a thermal expansion, where the Joule-Thomson coefficient $\mu_{JT}=\left.\frac{\partial T}{\partial P}\right)_H$  is the main quantity to discriminate between the cooling and heating regimes of the system. It is worth noting that when expanding a thermal system with a temperature $T$, the pressure always decreases yielding a negative sign to $\partial P$. In this context, we can consider two different regimes with respect  to the so-called the inversion temperature,  defined as the temperature $T_i$ at which the Joule-Thomson coefficient vanishes $\mu_{JT}(T_i)=0$ :  If $T<T_i\ \ (T>T_i)$, then the Joule-Thomson processus  cools (warms) the system with $\partial T<0$ and $\mu_{JT} >0$ ($ \partial T>0$ and $\mu_{JT}<0$) respectively. When the system temperature tends to $T_i$, its pressure is referred as the inversion pressure $P_i$, so defining a special point called the inversion point $(T_i, P_i)$ at which the cooling-heating transition occurs.

The outline of this work is as follows: In the next section, we review briefly the essential of the thermodynamic properties and stability of the charged-AdS black hole solution in $f(R)$ background. In section 3,  we study the {JT} expansion under constant mass, and derive the inversion temperature $T_i$   as well as the corresponding inversion point of such black hole. We also show when the cooling phase is changed to heating phase at a particular (inversion) pressure $P_i$. The last section is devoted to our conclusion.

  \section{  Thermodynamic of charged AdS black holes in $f(R)$ gravity background }  
  
 In this section, we briefly review the main features of the 
   four-dimensional charged AdS black hole
corresponding in the $R+f(R)$ gravity background
with a constant Ricci scalar curvature  \cite{fBH6}. The action is given by
\begin{eqnarray}
S=\int_{\mathcal{M}} d^{4}x\sqrt{-g}[R+f(R)-F_{\mu\nu}F^{\mu\nu}].
\label{action}
\end{eqnarray}
Here, $R$ denotes the Ricci scalar curvature while $f(R)$ is an arbitrary
function of $R$. In addition  $F_{\mu\nu}$ stands for electromagnetic field tensor given by
$F_{\mu\nu}=\partial_{\mu}A_{\nu}-\partial_{\nu}\partial_{\mu}$,
where $A_\mu$ is the  electromagnetic potential.
From the action \eqref{action}, the equations of
motion for gravitational field $g_{\mu\nu}$ and the gauge field
$A_{\mu}$ are,
\begin{eqnarray}
R_{\mu\nu}[1+f'(R)]-\frac{1}{2}g_{\mu\nu}[R+f(R)]+(g_{\mu\nu}\nabla^2-
\nabla_{\mu}\nabla_{\nu})f'(R)=T_{\mu\nu}, \label{Esteq}
\end{eqnarray}
\begin{eqnarray}
\partial_{\mu}(\sqrt{-g}F^{\mu\nu})=0.
\label{Emeq}
\end{eqnarray}

The analytic  solution of equation \eqref{Esteq} has been determined in \cite{fBH6}  when the Ricci scalar curvature is constant $R=R_0=const$, where in this simple case,  \eqref{Esteq} simplifies to:
\begin{eqnarray}
R_{\mu\nu}[1+f'(R_0)]-\frac{g_{\mu\nu}}{4}R_0[1+f'(R_0)]
=T_{\mu\nu}, \label{Esteq1}
\end{eqnarray}

The  charged static spherical black hole solution  of  \eqref{Esteq}  in in 4d gravity model takes the form
 \cite{fBH6}

 \begin{eqnarray}
ds^2=-N(r)dt^2+\frac{dr^2}{N(r)}+r^2(d\theta^2+\sin^2\theta
d\phi^2),\label{metric}
\end{eqnarray}

where the metric function $N(r)$ is given by,
\begin{eqnarray}
N(r)=1-\frac{2m}{r}+\frac{q^2}{br^2}-\frac{R_0}{12}r^2\label{metric1}
\end{eqnarray}
with $b=[1+f'(R_0)]$ while  the two parameters $m$ and $q$ are proportional to the black hole mass and the charge respectively
\cite{AS}
\begin{eqnarray}\label{mm}
M=m b, \quad  Q=\frac{q}{\sqrt{b}}.
\end{eqnarray}
In this background, the electric potential  $\Phi$ can be evaluated  as
\begin{eqnarray}
\Phi=\frac{\sqrt{b}q}{r_+}.
\end{eqnarray}
%

where the black hole event horizon $r_+$ denotes the largest root of the equation $N(r_+)=0$. At the event horizon, one can also derive the Hawking temperature as well as the  entropy of this kind of black hole \cite{fBH6,AS},\begin{eqnarray}
T=\frac{N'(r_+)}{4\pi}\bigg|_{r=r_+}=\frac{1}{4\pi
r_+}\bigg(1-\frac{q^2}{r_{+}^{2}b}
-\frac{R_0r_{+}^{2}}{4}\bigg),\label{Temp}
\end{eqnarray}
\begin{eqnarray}
S=\pi r_{+}^{2}b.\label{entropy}
\end{eqnarray}
Recalling the analogy between the cosmological constant  and the thermodynamics pressure, while its corresponding  conjugate quantity is identified to the volume \cite{sq1,Kastor,Cai:2001sn}, one can deduce the following relations, 
\begin{eqnarray}
P=-\frac{bR_0}{32\pi},\quad \text{ with }R_0=-\frac{12}{l^2}=-4\Lambda,\label{Pres}
\end{eqnarray}

and 
\begin{eqnarray}\label{volummme}
V=\frac{4\pi r^{3}_+}{3}. \label{Vo}
\end{eqnarray}

At this stage, it is straightforward to see that  the above black hole quantities satisfy to the following Smarr relation:
\begin{eqnarray}
M=2TS+\Phi Q-2PV. \label{Sm}
\end{eqnarray}
Furthermore, by taking into account the $F(R)$ corrections with a constant Ricci scalar, the first law of
thermodynamics is written as:
\begin{eqnarray}
d\bigg(\frac{M}{b}\bigg)=Td\bigg(\frac{S}{b}\bigg)+\bigg(\frac{\Phi}{b}\bigg)
dQ+Vd\bigg(\frac{P}{b}\bigg). \label{dfen}
\end{eqnarray}

At last, from  the equations of the Hawking temperature \eqref{Temp} and  the pressure
(\ref{Pres}) of such black hole, one can easily derive the corresponding equation of state, 
$P=P(T,r_+)$,
\begin{eqnarray}
P=\frac{bT}{2r_+}-\frac{b}{8\pi r_{+}^{2}}+\frac{q^2}{8\pi
r_{+}^{4}}.
\label{EOS}
\end{eqnarray}


It is worth noting here that 
the $F(R)$ background induced corrections to the pressure  (\ref{Pres}) and to the subsequent formulas can bring to light new possible feature which might be revealed through the phase transition structure of the charged-AdS black holes. Next section will be devoted to verify this proposal by means of Joule-Thomson expansion.

 \section{ Joule-Thompson expansion of Charged black hole in $f(R)$ gravity}  
 
Applying method similar to the one used in \cite{base}, we  consider Joule-Thomson expansion for charged-AdS black holes in $f(R)$ gravity. For a fixed charge the Joule-Thomson coefficient is given by, 
 \begin{equation}\label{38}
\mu=\left(\frac{\partial T}{\partial P}\right)_{M}=\frac{1}{C_{P}}\left[ T\left(\frac{\partial V}{\partial T}\right)_{P}-V\right],
 \end{equation}
besides, the equation of the state of such black hole is provided  in terms of thermodynamic volume  by substituting \eqref{volummme}  in \eqref{EOS}, the equation \eqref{Temp} transforms to:
 \begin{equation}\label{39}
T=\frac{1}{12V}\left(  \frac{12 \sqrt[3]{\frac{6}{\pi }} P V^{4/3}}{b}-4 Q^2+\left(\frac{6}{\pi }\right)^{2/3} V^{2/3} \right),
 \end{equation}
by using Eq.\eqref{39} this into the right hand side of Eq.\eqref{38}, one can derive the 
temperature   corresponding to a vanishing  Joule-Thomson coefficient, dubbed inversion temperature $T_i$:
 \begin{eqnarray}\label{40}
T_{i}=V\left(\frac{\partial T}{\partial V}\right)_P&=&\frac{1}{36 V}\left(\frac{12 \sqrt[3]{\frac{6}{\pi }} V^{4/3} P_i}{b}+12 Q^2-\left(\frac{6}{\pi }\right)^{2/3} V^{2/3}\right)\\ \nonumber
&=&\frac{1}{12 \pi  b r_+^3}\left(  -b r_+^2+3 b Q^2+8 \pi  r_+^4 P_i\right),
 \end{eqnarray}
$T_i$ can also be rewritten in terms of its corresponding pressure:
 \begin{eqnarray}\label{41}
T_{i}&=&\frac{1}{12 V}\left(  \frac{12 \sqrt[3]{\frac{6}{\pi }} V^{4/3} P_i}{b}-4 Q^2+\left(\frac{6}{\pi }\right)^{2/3} V^{2/3}    \right)\\ \nonumber
&=&\frac{1}{4 \pi  b r_+^3} \left(  b \left(r_+^2-Q^2\right)+8 \pi r_+^4 P_i \right).
 \end{eqnarray}
By subtracting Eq.  \eqref{40} form Eq. \eqref{41}, we get  the following polynomial equation,
 \begin{eqnarray}\label{poly}
-2 b r_+^2+3 b Q^2-8 \pi  r_+^4 P_i=0,
 \end{eqnarray}
 which  possesses four roots. Here, we only consider  the real positive  root given by,
 \begin{eqnarray}
r_+=\frac{\sqrt{\frac{\sqrt{b \left(b+24 \pi  Q^2 P_i\right)}}{\pi  P_i}-\frac{b}{\pi  P_i}}}{2 \sqrt{2}}.
 \end{eqnarray}
Once this root is substituted into Eq.\eqref{41}, the inversion temperature
becomes,
  \begin{eqnarray}\label{44}
T_i=\frac{\sqrt{P_i} \left(-\sqrt{b \left(b+24 \pi  Q^2 P_i\right)}+b+16 \pi  Q^2 P_i\right)}{\sqrt{2 \pi } \left(\sqrt{b
   \left(b+24 \pi  Q^2 P_i\right)}-b\right){}^{3/2}},
 \end{eqnarray}
so when  inversion pressure $P_i$ vanishes, the inversion temperature reaches its minimum at:
\begin{equation}
T_i^{min}=\frac{1}{6 \sqrt{6} \pi  Q}.
\end{equation}

Consequently, the critical temperature is just twice the value of the inversion temperature,
\begin{equation}
\frac{T_i^{min}}{T_c}=\frac{1}{2},
\end{equation}

in perfect agreement with the result of  \cite{jt1}.

\begin{figure}[!ht]
\begin{center}
\vspace{1cm}
\includegraphics[width=8cm]{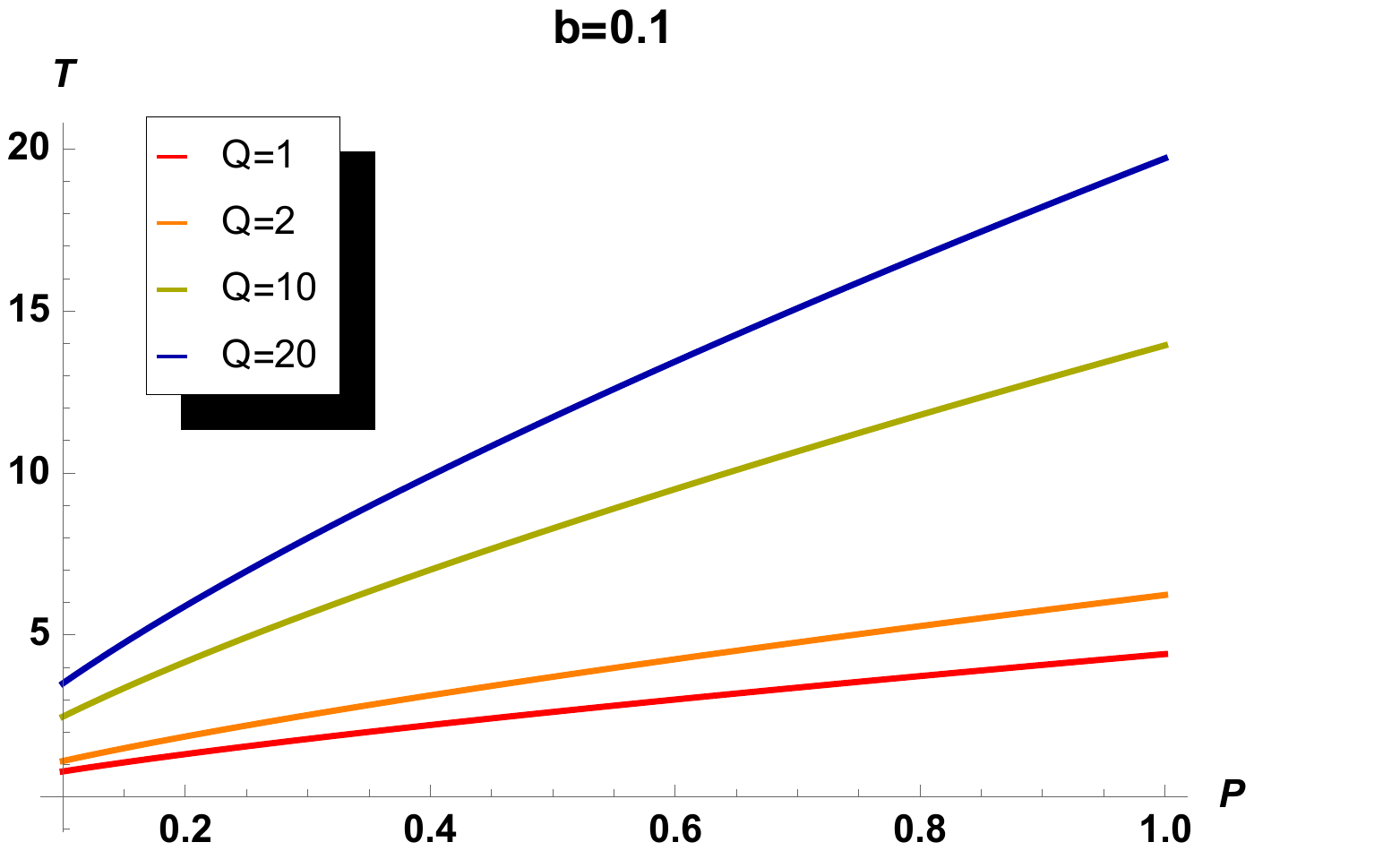}\vspace{1cm}\includegraphics[width=8cm]{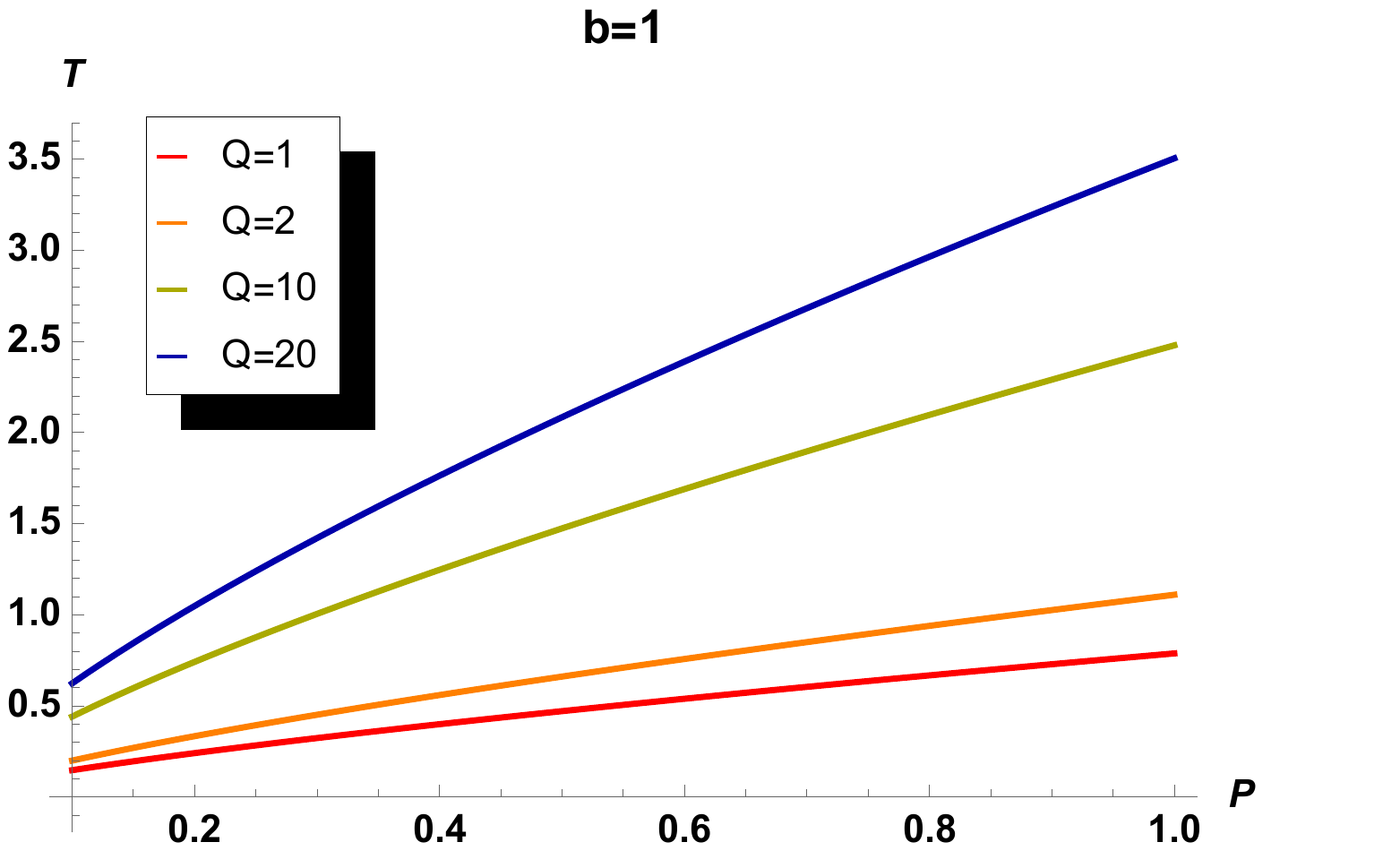}\vspace{1cm}
\includegraphics[width=8cm]{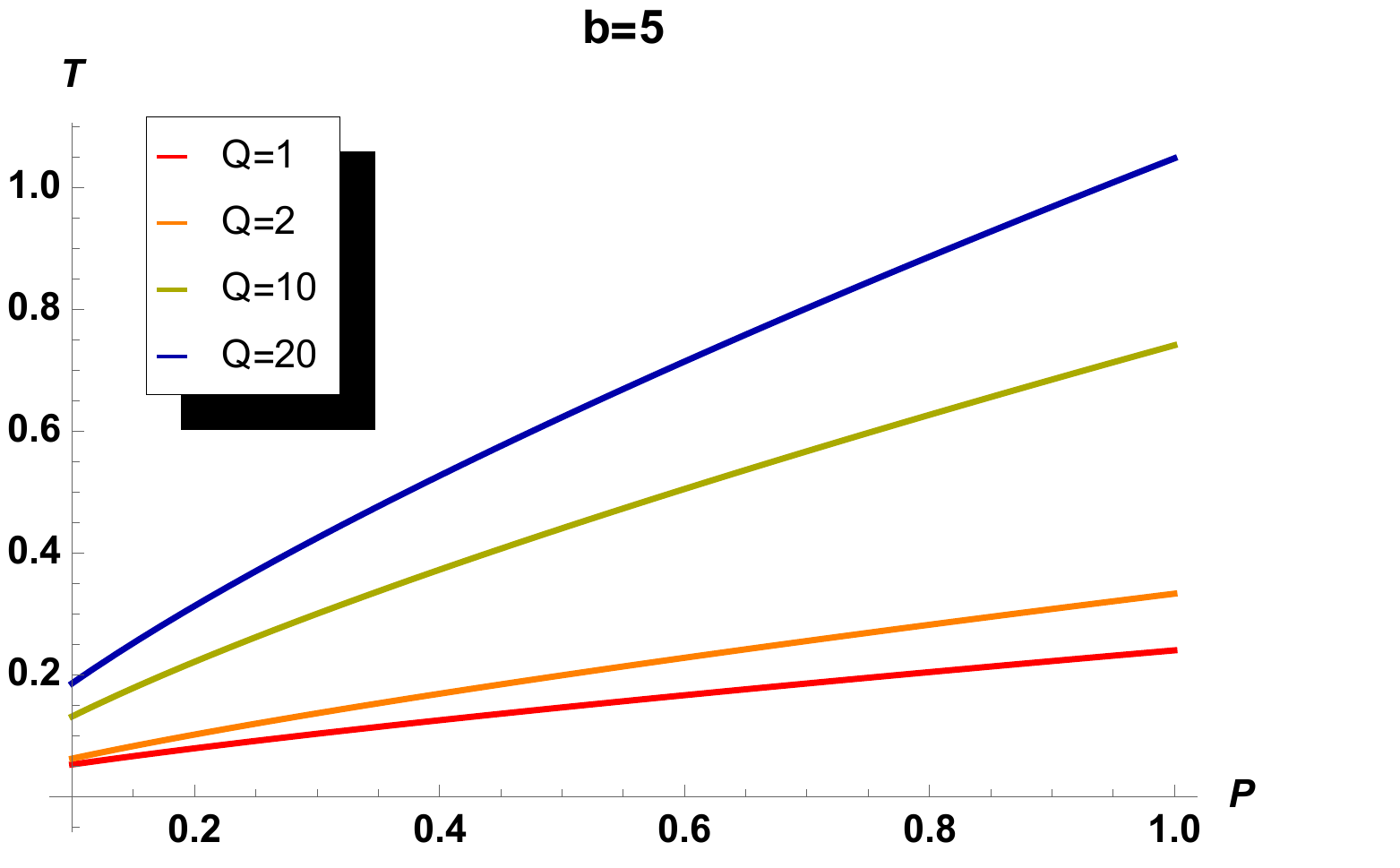}
\caption{Inversion curves. From bottom to top, the curves correspond to $Q = 1; 2; 10; 20$.}\label{figx}
\end{center}
\end{figure}

 In Fig.\ref{figx} we plot the inversion curves for charged AdS black hole for different values of charge $Q$. We can see  that,  in contrast to Van der Waals 
fluids, there is only a lower inversion curve which does not terminate at any point, since the expression inside
the square root of Eq.\eqref{44} is always positive. This feature has also been observed in \cite{jt1} with charged AdS black holes as well in \cite{jt2} with the rotating AdS black holes.

\begin{figure}[!ht]
\begin{center}
\includegraphics[width=8cm]{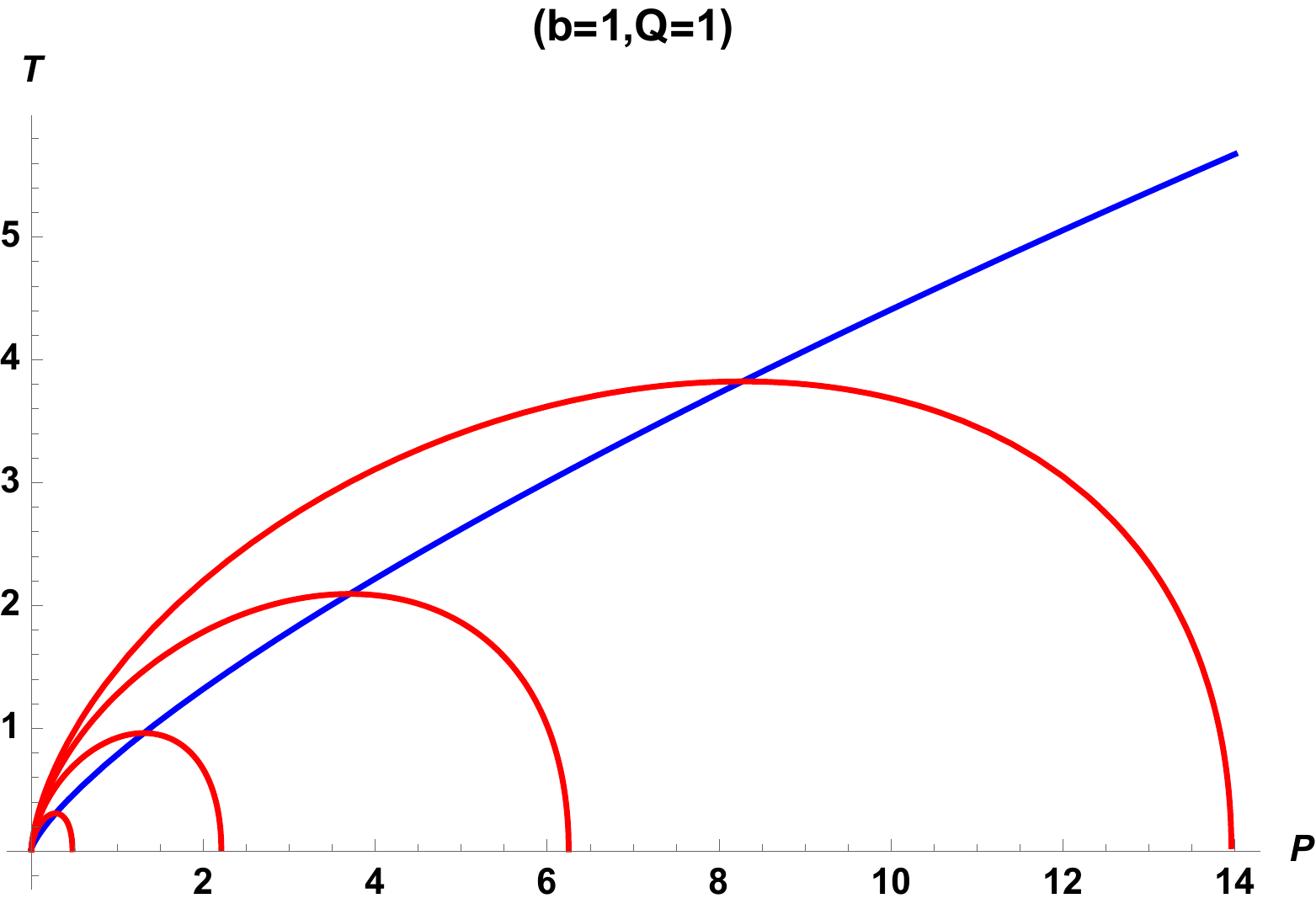}\vspace{1cm}\includegraphics[width=8cm]{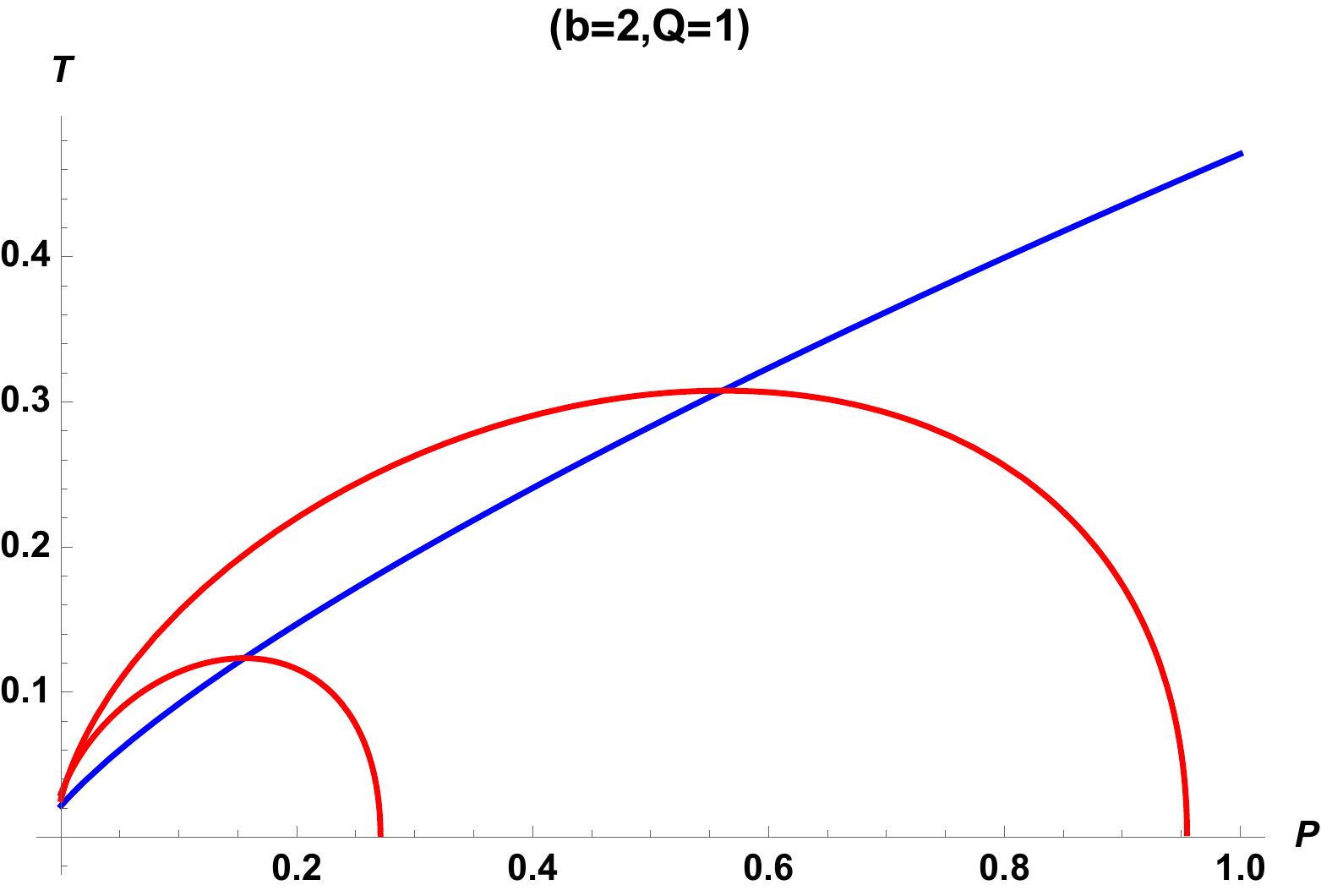}\vspace{1cm}
\caption{Inversion and isenthalpic curves for charged AdS black hole. From bottom to top, the isenthalpic curves correspond to
increasing values of M. Red and blue lines are isenthalpic and inversion curves, respectively. Our inputs are: Q = 1 , M = 1.5; 2; 2.5; 3 and b=1; 2. }\label{figy}
\end{center}
\end{figure}

 Next, thanks to Eq.\eqref{mm} and Eq.\eqref{EOS}, we also illustrates in Fig.\ref{figy} the isenthalpic curve corresponding to a constant mass in the $(T,P)$-diagram. 

 From Fig.\ref{figy}, we see that the inversion curves divide
the space $(T,P)$ into two separated regions: The  region above the inversion
curves corresponds to the cooling region, while the region under
the inversion curves corresponds to the heating one.  Note that one can discriminate between the cooling / heating regions just by checking the sign of the isenthalpic curves slope. The sign of the slope is
positive in the cooling region and it flip to negative in the heating region.  The cooling / heating phenomena  never  takes place on the inversion curve which plays the role of a separating boundary between the two regions.

\section{Conclusion}

In this paper, we have studied the  Joule-Thomson
expansion for RN-AdS black hole in $f(R)$ gravity  background
in the context of the extended phase space, where the cosmological constant is identified with the pressure. Here, the black hole mass is interpreted as an enthalpy,  so we assume that mass does not change during the Joule-Thomson expansion.  Our analysis has shown that the inversion curve always corresponds to the lower curve. This means
that black hole always cools above the inversion curve
during the expansion. At last, we have identified the cooling and heating
regions for different values of the parameter $b$ and the black hole mass $M$.

\vspace{10cm}
\newpage

\bibliographystyle{unsrt}
\bibliography{FRgravf}

\end{document}